\documentstyle[12pt,twoside,fleqn,psfig,espcrc1]{article}

\newcommand{\AmS}{{\protect\the\textfont2
  A\kern-.1667em\lower.5ex\hbox{M}\kern-.125emS}}
\title{
\vspace*{-4cm}
\begin{flushright}
DPNU-99-19\\
\vspace*{1.4cm}
\end{flushright}
Relativistic heavy ion collisions 
	--- Where are we now? Where do we go?}

\author{Masayuki Asakawa\\[1ex]
        Department of Physics, School of Science, Nagoya University, \\
	Nagoya 464-8602, Japan
        \thanks{This research was partly supported by Grant-in-Aid
	for Scientific Research	no. 10740112 of the Japanese Ministry
	of Education, Science, and Culture.}}

\begin{document}
\maketitle 
\begin{abstract}
Various aspects of the current status of ultrarelativistic heavy ion
collisions are reviewed. Perspectives of heavy ion physics in the
future are given as well.
\end{abstract}

\section{INTRODUCTION}

The construction of RHIC (Relativistic Heavy Ion Collider)
at BNL will be completed by the time these proceedings are out
and it will start accelerating gold nuclei at 100~GeV/A.
What is the purpose of ultrarelativistic heavy ion collisions? 
The reader may have dreams
such as i) study of high density matter, ii) production of the quark-gluon
plasma, iii) quasi-reproduction of the big bang, iv) understanding of
the history of the universe, and so forth. These are actually what will be
pursued at RHIC, and so these are not dreams any more. We, however, have
to turn our attention to the other side of reality as well:
i) the lifetime of the system created in ultrarelativistic
heavy ion collisions is very short, ii) the system is not
static, iii) observables are dirty, in other words, the interpretation
of observables is not straightforward, iv) there are
so many models that claim to describe the results of
ultrarelativistic heavy ion collisions successfully, and so on. 

Thus, in order to fully understand ultrarelativistic heavy ion collisions
and QCD, it is really necessary for theorists to attack the following
challenges, which were summarized by Matsui at Quark Matter 97
\cite{matsui98}:

{\noindent
(i) Compute, as best as one can, expected properties of dense matter and
its phase structure and make predictions for signals of new states of
matter.\\
(ii) Interpret the data from current fixed-target experiments and identify
signals of new physics, if any, from backgrounds of ``old physics''.\\
(iii) Describe formation and evolution of dense matter in nuclear collisions
and estimate physical conditions to be achieved at future collider
experiments.}

Since the first half of item (i) will be attacked by Yoshi\'{e}
\cite{yoshie},
I will concentrate on the second half of item (i) and
items (ii) and (iii). I will discuss mainly the recent results of
heavy ion collisions at CERN SPS at $E_{\rm lab} =$ 160 - 200~GeV/A.

\section{WHERE ARE WE?}

The system created in ultrarelativistic heavy ion collisions 
is not static. Even if hot/dense hadron matter or quark-gluon plasma
is created, it cannot last for long time. The system immediately
starts to expand and cool down. Even if the quark-gluon plasma phase
is produced initially, it is soon converted back to the hadron matter,
which freezes out in a time scale of a few tens of fm.
A schematic diagram of the time evolution in ultrarelativistic heavy ion
collisions is shown below.
\vspace*{0.0cm}
\begin{figure}[hbt]
\hspace*{3.12cm}
\psfig{file=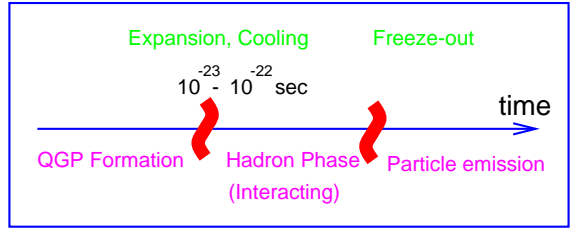,height=1.2in,angle=0}
\vspace*{-0.5cm}
\caption{Schematic time evolution expected in ultrarelativistic
heavy ion collisions.}
\vspace*{-0.2cm}
\end{figure}

This kind of explanation is often seen in the literature.
However, the reader may wonder if this is indeed what is realized in
ultrarelativistic heavy ion collisions. Without evidence, the whole picture
is a mere conjecture. In the following, I will answer this question
with recent experimental results and theoretical inference, and
support this picture.

\subsection{Are hot systems ever produced in ultrarelativistic
heavy ion collisions?}

Probably the reader is most interested in the possibility of the production
of the quark-gluon plasma in ultrarelativistic heavy ion collisions.
Nevertheless, I would like to begin with a more fundamental question;
are there thermalization and collective motion in ultrarelativistic
heavy ion collisions? It is because the phase transition to the
quark-gluon plasma is a result of interaction. If no bulk interacting
system is created, no phase transition takes place. It is logically
possible that two nuclei collide on each other and go through each
other without leaving a region with high energy density.
However, recent experimental data tell us that approximately
thermalized systems are indeed created in
ultrarelativistic heavy ion collisions and that the systems go
through collective expansion. The evidence includes the result of
direct measurement of flow \cite{e877},
measurement of the interference of
identical particles \cite{na44}, and so on. Here I will discuss the
transverse mass, $m_T$, distribution of final state hadrons \cite{na49}.

The transverse mass, $m_T$, is defined by
$m_T^2 = m^2 + p_T^2$, where $m$ is the mass of the particle and
$p_T$ is its transverse momentum. The beam direction is defined
as the longitudinal axis. The $m_T$ distribution of final
state hadron, $i$, at mid rapidity is well-approximated
by the following form:
\begin{equation}
\frac{1}{m_T}\frac{dN}{dm_T} \propto \exp
\left ( - \frac{m_T}{T_i} \right ).
\end{equation}
$T_i$ has been called temperature or slope parameter.
It is known that $T_i$ is a function of particle mass and
well-approximated as
\begin{equation}\label{slope}
T_i = a + bm_i,
\end{equation}
where $a$ and $b$ are constants dependent on colliding nuclei,
collision energy, and event class such as central or peripheral.
This mass dependence of $T_i$ has the following simple interpretation:
$a = T_f$ and $b=\langle v^2 \rangle$, where $T_f$ is the temperature of
the system at freeze-out and $v$ is the flow velocity at freeze-out.
This relation is derived by assuming that all particles are
locally in thermal equilibrium (not necessarily in chemical equilibrium)
and collectively expanding at freeze-out and that the freeze-out
time is independent of particle species. Thus, the experimental
observation, Eq. (\ref{slope}), is not inconsistent with the
formation of hot thermalized system.

Note that, as I emphasized above, this is merely one of the data that
support the formation of hot systems. If this were the only evidence,
the formation of hot systems cannot be concluded, since even in
high energy $pp$ collisions the $m_T$ distribution is exponential
\cite{heinz}.
By combining various observables, the state of the system is deduced.
This is the heavy ion way of inference.

\subsection{What kind of matter is created?}

We have learned that some kind of interacting matter is created,
at least transiently, in ultrarelativistic heavy ion collisions.
The next question is what kind of matter is created.
One of the ways to get clues to it is to measure $J/\psi$ yield.

As is well known, the suppression of $J/\psi$ yield was originally
proposed as a signature of the formation of the quark-gluon plasma
\cite{ms86}.
The idea was that if the quark-gluon plasma is created, $J/\psi$
cannot form because the potential between a $c\bar{c}$ pair is
Debye screened. However, this is not the only process that
suppresses $J/\psi$ yield. Processes such as
$J/\psi + N \rightarrow D\bar{D}N$,
can contribute to $J/\psi$ suppression. Thus, $J/\psi$ suppression
is not necessarily a signature of the quark-gluon plasma by itself.
Until ${\rm Pb}{\rm Pb}$ experiments began at CERN,
$J/\psi$ data in pA and AB collisions had been successfully explained by
the above process, i.e., $J/\psi$ absorption by the nucleon.
In ${\rm Pb}{\rm Pb}$ collisions at $E_{\rm lab} = 160$ GeV/A,
however, it was found that as the transverse energy $E_T$ of
final state hadrons increases, the $J/\psi$ yield, more strictly
$B_{\mu\mu}\sigma(J/\psi)/\sigma({\rm Drell-Yan})$, drops
suddenly at an $E_T$ \cite{na50}. Hadronic scenarios failed to explain
this behavior \cite{vogt}. However, the nature of this sudden drop is
not clear yet, although there are a lot of attempts to explain
the behavior \cite{geigerch}. In particular, I remark that
even if the phase transition from the hadronic phase to
the quark matter is of first order, it cannot lead to
the sudden drop of the $J/\psi$ yield in such a naive
way as discussed in \cite{satz}. It is because
the energy density jumps at a first order phase transition.
Thus, as $E_T$ is increased, part of the system begins
to become the quark matter at $E_{T1}$,
and the portion of the quark matter increases gradually.
When the whole volume has become the quark matter, $E_T$ must have
become much larger than $E_{T1}$. As a result, the sudden drop
in the $J/\psi$ yield does not correspond to the sudden formation
of the quark-gluon plasma at a first order phase transition.

\subsection{How is the matter being excited?}

We have seen that some new form of matter appears to
be created in ultrarelativistic heavy ion collisions.
Then, how is the matter excited quantum mechanically?
Dileptons are suitable probes to study this problem.
Leptons do not interact strongly and can be considered
almost penetrative. Observed dileptons carry the information
of the early hot/dense stage as well as later stages, and some dileptons
are produced even after freeze-out, for instance, by the decay
of vector mesons. Accordingly, if vector mesons are modified,
it is expected to be observed with dileptons, but not with hadrons.

The first evidence of hadron modification was brought by CERES
Collaboration at CERN SPS \cite{ceres}. They first measured dilepton and
meson yields in $p{\rm Be}$ and $p{\rm Au}$ collisions, and found that
the dilepton spectra in those reactions are explained solely by the decay
of final state mesons. This means that no long-lived fireball is created
in $p{\rm Be}$ or $p{\rm Au}$ collisions. They have, however,
shown that the dilepton spectra in ${\rm S}{\rm U}$ collisions
at $E_{\rm lab} = 200$ GeV/A cannot be explained only by the decay
of final state mesons. Later, it was found that the data cannot be
reproduced without taking account of hadron modification in the hot
phase created in the collisions \cite{hung,ko}.

Thus, hadrons are modified in medium. Two scenarios are often
compared: mass shift and collisional broadening. The two scenarios
are often considered two different scenarios. This is, however, not
the case. First, I point out that the term, mass shift, is quite
confusing. What is observed with dileptons is not the pole masses of
vector mesons but the spectral function. In general, for 
the vector Heisenberg operator $J_\mu(\vec{x},t)$ the polarization tensor
$\Pi_{\mu\nu}(q_0,\vec{q})$ is defined by
\begin{equation}
\Pi_{\mu\nu} = i \!\int\! d^4 x e^{iqx} \langle
TJ_\mu (\vec{x},t) J_\nu^\dag (\vec{0},0)\rangle_T ,
\end{equation}
where $\langle\cdots\rangle_T$ indicates thermal average at $T$.
For simplicity, let us set $\vec{q} = \vec{0}$.
The spectral function $\rho(q_0 )$ is related to
the polarization tensor,
\begin{equation}
\rho(q_0 ) \propto {\rm Im} \frac{1}{q_0^2}
\frac{\Pi_\mu^\mu (q_0)}{\tanh(\beta q_0/2)}, \quad \beta = \frac{1}{T}.
\end{equation}
The physical significance of the spectral function is that
the dilepton production rate at $T$,
$(dN_{\ell\bar{\ell}}/d^4 x d^4 q )_{T}$,
is related to the spectral function without approximation as follows
\cite{weldon,gk}:
\begin{equation}\label{dilepton}
\left (\frac{dN_{\ell\bar{\ell}}}{d^4 x d^4 q }\right )_{T}
\propto \frac{e^{\beta q_0} + 1}{(e^{\beta q_0} - 1)^2}
\rho(q_0).
\end{equation}
This formula is exact, independent of in what phase the system is.
There is some confusion on the meaning of dilepton spectra.
Some authors argue that observed dilepton spectra are
different from those theorists calculate because final state
interactions modify theoretically calculated spectra.
This statement stems from the misunderstanding that
masses calculated by theorists are the masses of quasi-particles at
$T$ obtained by diagonalizing the Hamiltonian. If this were the case,
the effect of final state interactions
may change theoretical predictions substantially and 
should be taken into account. However, the formula,
Eq. (\ref{dilepton}), is exact, and so no further
interface between experiments and theories
is needed except purely experimental corrections.

Collisional broadening is a universal phenomenon; it
appears wherever collisions take place \cite{gallagher}.
Generally, hadronic effective theories calculate this part of
hadron modification. The other type of hadron modification,
mass shift or global shift of peaks in the spectral function,
is special to QCD. In the QCD sum rules, this is expressed as follows
\cite{hatsuda}.
The operator product expansion side of the polarization function
is given by
\begin{equation}
\frac{i}{3Q^2}\!\int\! d^4 s e^{i\omega t}\langle
TJ_\mu (\vec{x},t)J^{\mu\dag}(\vec{0},0)\rangle _T
= -C_0 \log |Q^2 | + \sum_{n=1}^{\infty}\frac{C_n}{Q^{2n}},
\end{equation}
where $Q^2 = -\omega^2$, $C_i$'s are condensates, and
$\vec{q} = \vec{0}$ has been assumed as before.
The condensates are related to the spectral function by
the following exact sum rules:
\begin{equation}\label{sumrule}
\int_{0}^{s_0}\!\rho(\sqrt{s})s^n ds = \frac{C_0}{n+1}s_0^{n+1}
+(-1)^n C_{n+1}, \quad n\geq 0 ,
\end{equation}
where $s_0$ is the perturbative QCD threshold.
At finite temperature or density, the condensates change because of
partial chiral restoration...etc.. The change is reflected to
global shift of peak structure in the spectral function at
finite temperature or density through the exact sum rules, (\ref{sumrule}),
in addition to the trivial collisional broadening
\cite{ak93}.
Thus, the two scenarios for hadron modification are not exclusive
with each other. In QCD both mechanisms are indeed at work
and should be taken into account. In particular, the sum rules,
(\ref{sumrule}), should be satisfied by every effective model as well.

\subsection{Possibility of non-equilibrium states?}

Heavy ion reactions take place within a finite time. The typical
time scale is of the same order as that of strong interaction and the
process does not necessarily proceed adiabatically. Therefore, there
is plenty of potential room for non-equilibrium phenomena.
This non-equilibration is not simply limited to phase space,
but also expected in chiral space. 

One of such possibilities is disoriented chiral condensate (DCC).
Rajagopal and Wilczek proposed
a mechanism to create DCC domains called `quench mechanism'
\cite{rw93}.
I use, as a model Lagrangian,
the linear sigma model defined by
\begin{equation}\label{lsigma}
{\cal L} = \frac{1}{2}\partial_{\mu}\phi_i \partial^{\mu}\phi_i
-\frac{\lambda}{4}(\phi^2 -v^2)^2 + H\sigma,
\end{equation}
where $\phi_i \equiv (\sigma, \mbox{\boldmath $\pi$})$ stands for a vector
in internal space; $H\sigma$ is an explicit chiral symmetry breaking term
due to the finite quark masses.

The original idea of the quench mechanism for DCC formation is
summarized as follows. First, chiral symmetry is restored in the
central region in particle or nuclear collisions. Then, the chiral
fields are assumed to decouple quickly from the heat bath. The chiral
fields are thus left around the origin in chiral space, i.e.,
$\phi^2 \sim 0$, while the effective potential has returned to
its zero temperature form. As a result, the chiral
fields are left at the top of the so-called `Mexican hat' effective
potential and according to its initial condition at each spatial point,
the chiral field $\phi$ rolls down the slope of the effective potential
in an arbitrary direction toward the bottom of the potential in chiral space.
If in a certain spatial region the chiral field collectively rolls
down in approximately the same direction and acquires an expectation
value different from that in the vacuum, it will result in a DCC domain
in coordinate space. This is schematically shown in Fig. 2. 

\begin{figure}[hbt]
\begin{minipage}[t]{75mm}
\psfig{file=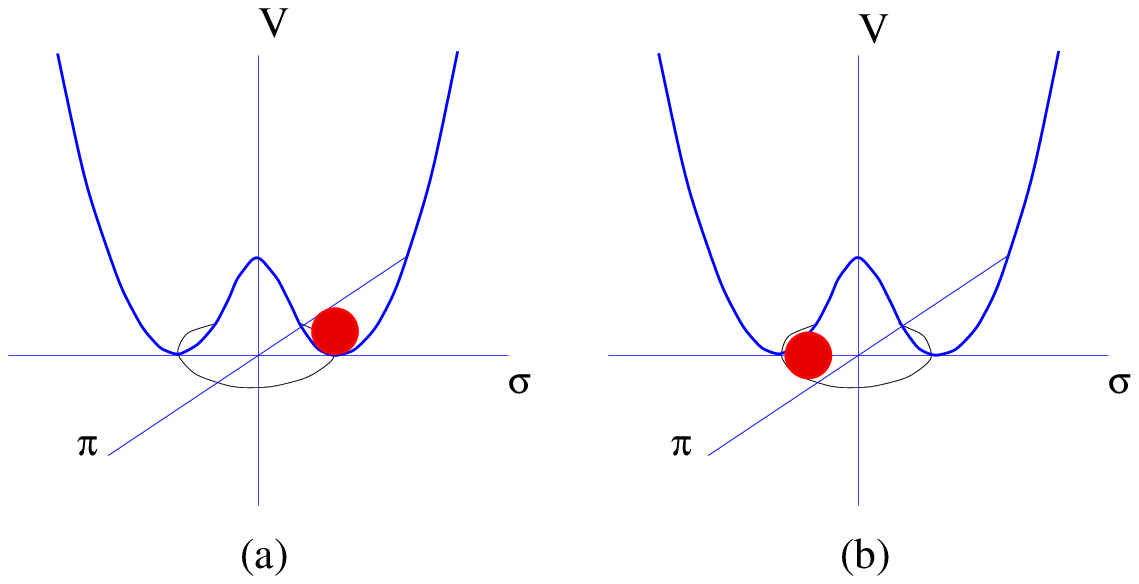,height=1.5in,angle=0}
\vspace*{-0.5cm}
\caption{Configuration of the chiral field (a) in the vacuum
and (b) in a DCC state.}
\end{minipage}
\hspace*{0.7cm}
\begin{minipage}[t]{75mm}
\psfig{file=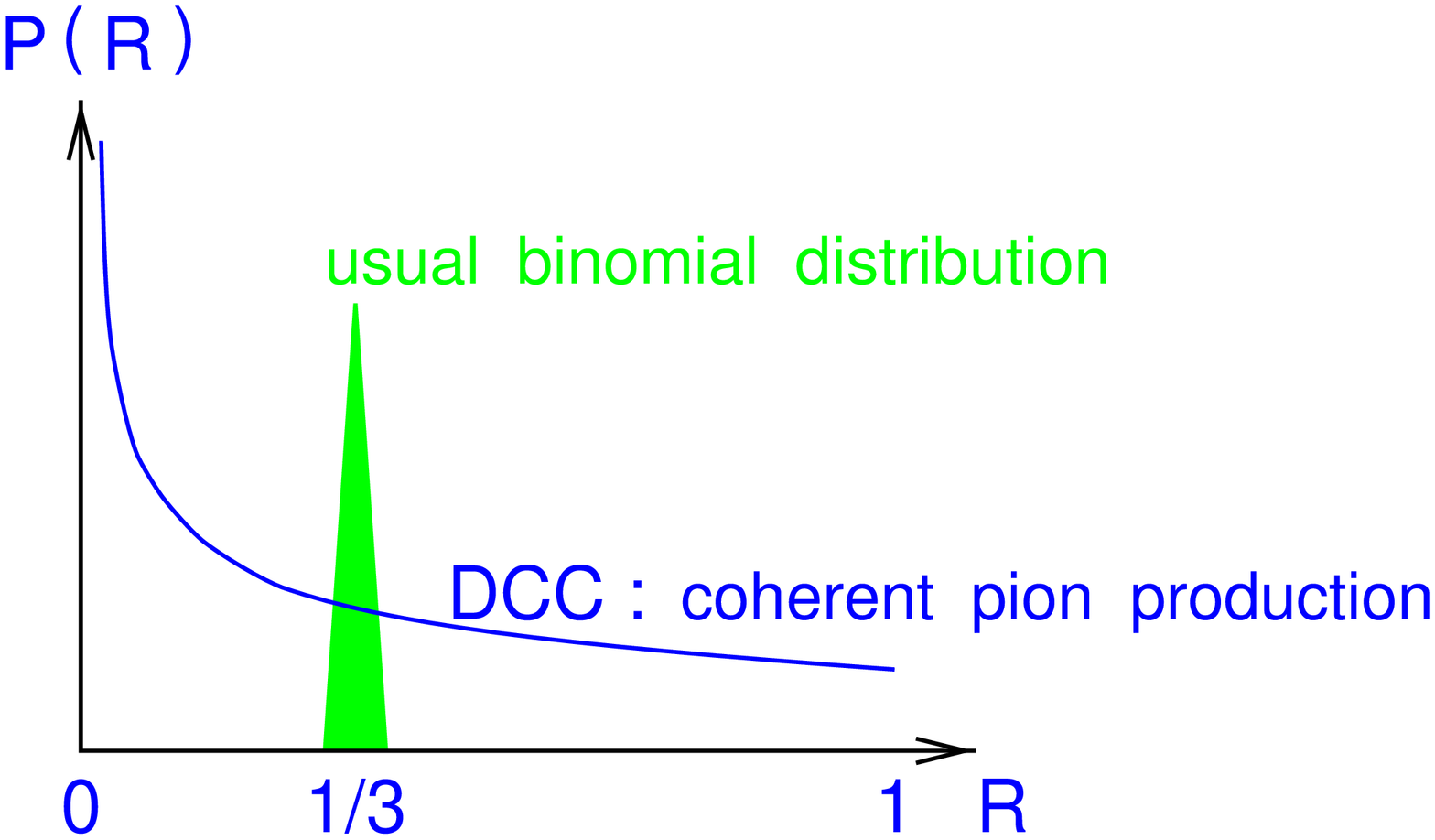,height=1.5in,angle=-90}
\vspace*{-0.5cm}
\caption{Distribution probability of $R$ defined by Eq. (\ref{rdef}).}
\end{minipage}
\vspace*{-0.2cm}
\end{figure}

Actually, the above explanation tells us why we can expect
non-equilibrium states in chiral space, but does not tell us what is the
mechanism to create `domains', i.e., large scale structure. It is
the amplification of low momentum modes caused by the mode instability
in the Mexican hat effective potential. The equation of motion
for the momentum ${\bf k}$ component of the pion field
$\pi^i$ is given in the mean field approximation,
\begin{equation}\label{mfequation0}
\frac{d^2 \mbox{\boldmath $\pi$}_{\bf k} }{dt^2} = 
[\lambda(v^2 - \langle\phi^2
\rangle ) -k^2 ] \mbox{\boldmath $\pi$}_{\bf k},
\end{equation}
where the fluctuations of the chiral fields were neglected and
$\langle\phi^2 \rangle$ is the average of $\phi^2 = \phi^i \phi^i$.
In this approximation, modes with $k^2 < v^2 - \langle\phi^2 \rangle$
grow exponentially, while high momentum modes do not. As a result,
a large scale structure is expected to emerge.

If a DCC domain is created and all final state pions are emitted
from the single domain, it can be shown, using the remaining
$O(3)$ invariance, that the distribution probability of a quantity
$R$ defined by 
\begin{equation}\label{rdef}
R=\frac{N_{\pi_0}}{N_{\pi_0}+N_{\pi^+}+N_{\pi^-}},
\end{equation}
where $N_{\pi_0}$, $N_{\pi^+}$, and $N_{\pi^-}$
are the numbers of the final state $\pi_0$, $\pi^+$, and $\pi^-$,
respectively, takes the following form \cite{blaizot}:
\begin{equation}\label{poisson}
P(R)=\frac{1}{2\sqrt{R}}.
\end{equation}
Note that $P(R)$ diverges at $R=0$, but that the average of
$R$ takes a finite value, 1/3. This probability distribution is
quite in contrast to that expected for normal incoherent emission,
$P(R) \sim \delta(R-1/3)$ (See Fig. 3).
This behavior is valid only for low $p_T$ pions,
because DCC formation is caused by amplification
of low momentum pion modes.

Eq. (\ref{mfequation0}) tells us that it is necessary to get the system
cooled fast enough to have (i) spontaneously symmetry broken effective
potential and (ii) discordance between the minimum point of
the spontaneously chirally broken effective potential and
the position of the mean field in order to have large and
discernible domains.

In central ultrarelativistic heavy ion collisions, the initial fluctuation
is expected to be large and the typical time scale for the expansion
of the system is also large. Thus, the system is likely to maintain
equilibration at least in chiral space. The above conjecture has been
also confirmed numerically by our group \cite{ahw95},
assuming the Bjorken scaling in the longitudinal
direction \cite{bjorken83}.
This leads to the following conclusion: central
ultrarelativistic heavy ion collisions are not the best place
to look for DCC formation, contrary to the general expectation.
It is not-so-central ultrarelativistic heavy ion collisions
that should be collected to look for DCC formation.
I also note that in not-so-central ultrarelativistic heavy ion
collisions the axial anomaly is expected to bring about DCC domain
formation aligned in real space, i.e., one above the reaction place
and one below it, but misaligned in chiral space \cite{amm}.
This conjecture can be tested by utilizing the technique to
determine the reaction plane, which has been developed
in flow analysis.

At the moment, only one heavy ion experiment has reported the result
of a DCC search \cite{wa98}. Their result was negative.
However, they used only central collisions and did not imposed
a cut on $p_T$. As I discussed above, it is necessary to select
non-central but non-peripheral collisions and low $p_T$ pions
in DCC hunts. Until experiments with such cuts are carried out,
the possibility of DCC formation in ultrarelativistic heavy ion
collisions will remain an open question.

\section{WHERE DO WE GO?}

RHIC is coming late this year (1999) at $\sqrt{s} = 200$ GeV/A
(A = $^{197}\!{\rm Au}$). Hopefully, LHC will come early next century. 
While physical observables and methodology will remain essentially
the same, energy will be much higher, momentum and invariant mass
resolution will be much better (the $e^+ e^-$ invariant mass resolution
is about 20 - 30 times better at 1~GeV than the current CERES setup
at CERN SPS), and acceptance will be much larger.
The quark-gluon plasma, whose formation is not decisive at the moment,
will be created with much larger probability. Since the decrease of
the temperature is slowed near a phase transition irrespective of
the order of the transition, if the phase transition occurs,
it will be possible to separate dileptons from the (almost) constant
temperature period near the critical temperature as a secondary peak
or shoulder structure in the invariant mass distribution \cite{ak}.

The system created in ultrarelativistic heavy ion collisions is
very complex. In order to obtain profound understanding of the small
short-lived non-static and potentially non-equilibrated system,
it is mandatory to compare as many observables as possible and find
correlations among them. The comparison should not be limited to
event classes at a fixed collision energy, but should be among events
at different collision energies. So far, there has been
a large gap in collision energies of heavy ion experiments
between $E_{\rm lab}=\,\sim\!$ 12~GeV/A
(BNL AGS) and $\sim\!$ 160~GeV/A (CERN SPS). The gap between
CERN SPS and RHIC is considerably large as well.
From this viewpoint, the approval and inauguration of the JHF project,
which is capable of accelerating heavy ions at $E_{\rm lab} = $ 50~GeV/A
and lower, are eagerly awaited. It is also desirable to run RHIC at lower
energies.
\vspace*{0.0cm}
\begin{figure}[hbt]
\hspace*{2.5cm}
\psfig{file=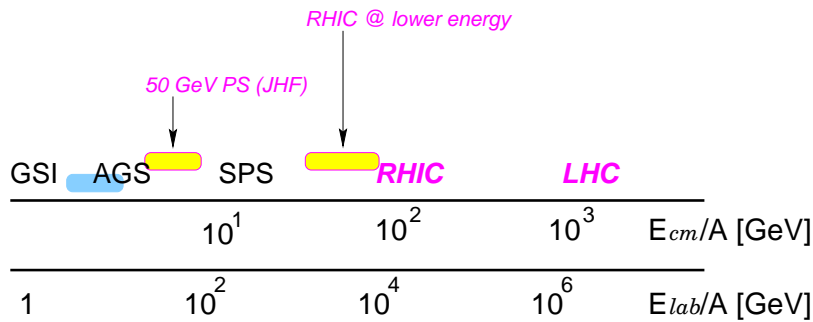,height=1.6in,angle=0}
\vspace*{-0.5cm}
\caption{Energies of past, present,
and future high energy heavy ion machines.}
\vspace*{-0.2cm}
\end{figure}

Finally, I make comments on event generators. Quite a number of
event generators exist in the market \cite{pang}.
Most of them are classical, although Pauli blocking is
generally taken into account. This is acceptable as a first
step. However, it is not acceptable that most models do not
include quarks or gluons and that many models do not include
secondary collisions. This is completely contrary to the
general expectation that RHIC physics is described by quark
and gluon degrees of freedom and the fact that phase transition takes
place owing to interaction among quarks and gluons. In this respect,
perhaps the most ambitious attempt was Parton Cascade (or VNI)
\cite{kkg} by Klaus Kinder Geiger, although it is not
without serious conceptual defects yet. He tragically perished
in the Swiss Air Flight 111 crash on September 2, 1998, but
I hope that this is not the end. For better understanding of
the dynamical features of RHIC and LHC physics, the direction initiated
by him should be taken over and further pursued.

\end{document}